 \documentclass[final,3p,times,twocolumn]{elsarticle}

\usepackage[inkscapeformat=png]{svg}
\usepackage[format=plain]{caption}
\usepackage{subcaption}
\usepackage{braket}
\usepackage{float}
\usepackage{amsmath,mathtools}
\usepackage[colorlinks]{hyperref}
\usepackage{flushend}
\usepackage{dblfloatfix}

\usepackage{amssymb}

\newcommand{\barium}{$\prescript{136}{}{\text{Ba}}$\space}
\newcommand{\xenon}{$\prescript{136}{}{\text{Xe}}$\space}

\newcommand{\bariumium}{$\prescript{136}{}{\text{Ba}^+}$\space}
\newcommand{\ovbb}{$0\nu\beta\beta$\space}
\newcommand{\tuvbb}{$2\nu\beta\beta$\space}

\journal{arXiv}

\begin{document}

\begin{frontmatter}



\title{\vspace*{-0.5cm}`Searching for a needle in a haystack;' A Ba-tagging approach for an upgraded nEXO experiment}

\author[1]{H. Rasiwala\corref{cor1}}
\cortext[cor1]{Corresponding author.}
\ead{hussain.rasiwala@mail.mcgill.ca}
\author[1]{ K. Murray}
\author[2,11]{ Y. Lan}
\author[1]{ C. Chambers\fnref{a}}
\author[2,10]{ M. Cvitan}
\author[1,2]{ T. Brunner}
\author[5]{ R. Collister}
\author[6]{ T. Daniels}
\author[5]{ R. Elmansali}
\author[7]{ W. Fairbank}
\author[5]{ R. Gornea}
\author[8]{ G. Gratta}
\author[5]{ T. Koffas}
\author[2]{ A.A. Kwiatkowski}
\author[3,4]{ K.G. Leach}
\author[2,10]{ A. Lennarz}
\author[2]{ C. Malbrunot}
\author[1,2]{ D. Ray}
\author[5]{ R. Shaikh}
\author[9]{ L. Yang}
\author[]{ for the nEXO Ba-tagging group}

\fntext[a]{Now at TRIUMF, Vancouver, Canada}

\address[1]{Department of Physics, McGill University, Montreal, QC, H3A 0G4, Canada}
\address[2]{TRIUMF, Vancouver, BC, V6T 2A3, Canada}
\address[11]{Department of Physics, The University of British Columbia, Vancouver, BC, V6T 1Z4, Canada}
\address[10]{Department of Physics and Astronomy, McMaster University, Hamilton, ON, L8S 4L8, Canada}
\address[5]{Department of Physics, Carleton University, Ottawa, ON, K1S 5B6, Canada}
\address[6]{Department of Physics and Physical Oceanography, University of North Carolina at Wilmington, Wilmington, NC, 28403, USA}
\address[7]{Physics Department, Colorado State University, Fort Collins, CO, 80523, USA}
\address[8]{Physics Department, Stanford University, Stanford, CA, 94305, USA}
\address[3]{Department of Physics, Colorado School of Mines, Golden, CO, 80401, USA}
\address[4]{Facility for Rare Isotope Beams, Michigan State University, East Lansing, MI, 48824, USA}
\address[9]{Department of Physics, University of California San Diego, La Jolla, CA, 92093, USA}

\begin{abstract}
nEXO is a proposed experiment that will search for neutrinoless double-beta decay (0$\nu\beta\beta$) in 5-tonnes of liquid xenon (LXe), isotopically enriched in \xenon\hspace{-3pt}. A technique called Ba-tagging is being developed as a potential future upgrade for nEXO to detect the \xenon double-beta decay daughter isotope, \barium\hspace{-3pt}. An efficient Ba-tagging technique has the potential to boost nEXO's 0$\nu\beta\beta$ sensitivity by essentially suppressing non-double-beta decay background events. A conceptual approach for the extraction from the detector volume, trapping, and identification of a single Ba ion from 5 tonnes of LXe is presented, along with initial results from the commissioning of one of its subsystems, a quadrupole mass filter.
\end{abstract}

\begin{keyword}
Barium-tagging \sep nEXO \sep Neutrinoless double-beta decay. 
\end{keyword}

\end{frontmatter}


\section{Introduction}
\label{intro}

Cosmological observations provide compelling evidence that the Universe is dominated by matter, which contradicts the expectation that the Big Bang produced similar amounts of matter and anti-matter \cite{matter_asym}. Leptogenesis is a theory that explains this asymmetry but requires the violation of lepton number in weak decays, which is so far a conserved quantity in the Standard Model of particle physics (SM) \cite{leptogenesis}. Following the discovery of neutrino oscillations, neutrinos have emerged as a promising candidate for the observation of lepton number violation, as there is now a possibility that neutrinos are their own antiparticles, referred to as Majorana particle \cite{majorana}. One of the most effective ways to probe the Majorana nature of the neutrino is through the search for neutrinoless double-beta decay (\ovbb\hspace{-3pt}).

\begin{figure*}[b]
    \centering
    \includegraphics[width=0.75\textwidth]{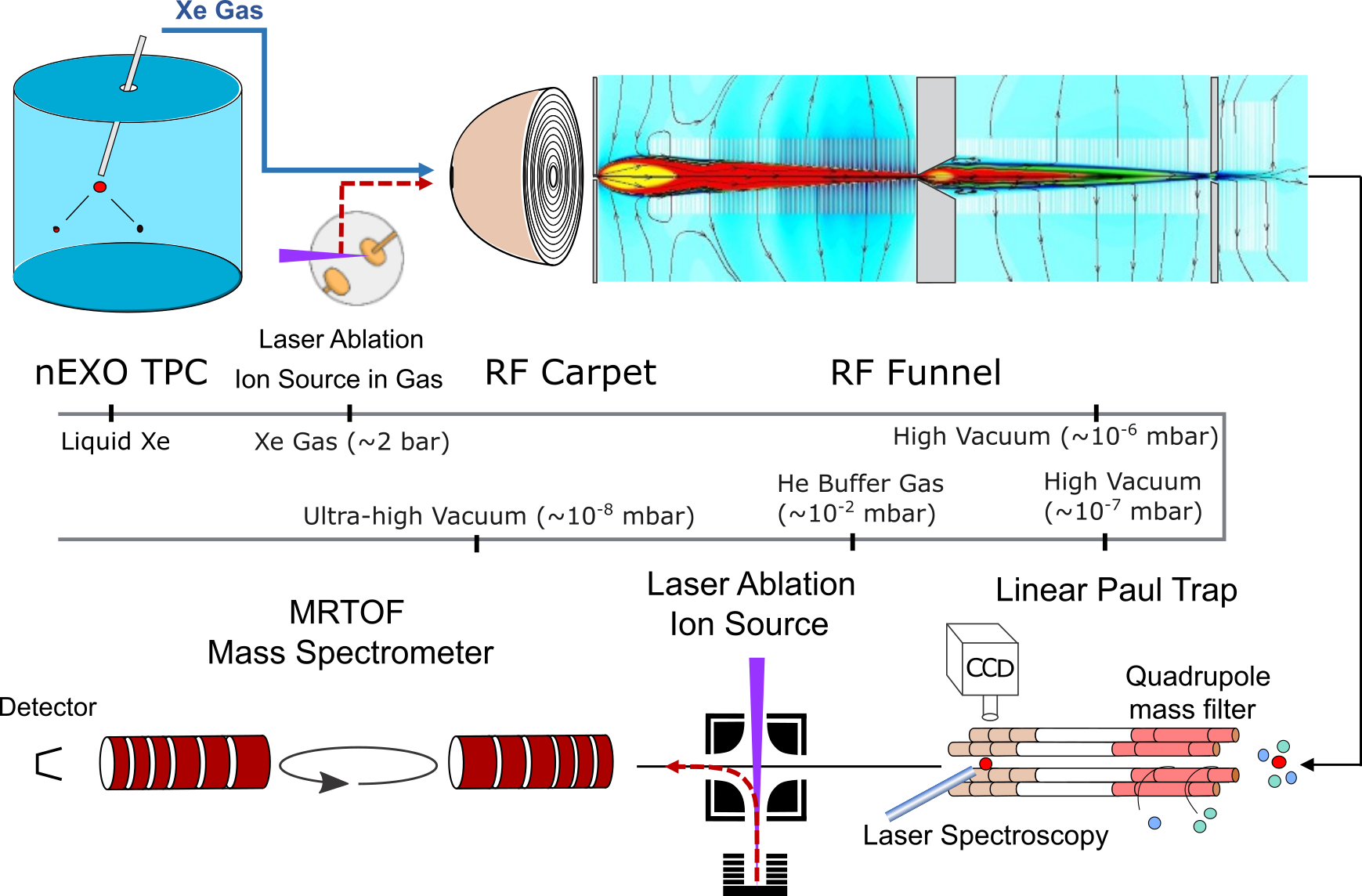}
    \caption[Schematic of the entire Ba-tagging assembly.]{Schematic of the entire Ba-tagging scheme. The \bariumium is extracted from the TPC along with LXe. Following the phase transition to Xe gas, an RF carpet facilitates \bariumium injection into the RF funnel, which transports the \bariumium from high-pressure Xe (2-3 bar) to high vacuum ($<10^{-6}$ mbar). It is then transferred to a linear Paul trap for filtering, cooling, and trapping. Here the Ba ion is identified using laser fluorescence spectroscopy and then transferred to a mass spectrometer to determine its mass.}
    \label{fig:batag}
\end{figure*}

\ovbb is a Standard Model (SM) forbidden decay which, if observed, would violate the conservation of lepton number and provide insight into the mass-scale and mass-ordering of neutrinos along with experimental proof of their Majorana nature. Double-beta decay (\tuvbb\hspace{-3pt}) is a second-order weak nuclear decay that involves the simultaneous decay of two neutrons to two protons, two electrons, and two electron-type antineutrinos. This is a rare type of nuclear decay seen only in a few even-even nuclei \cite{dolinski}, where single beta decay is energetically forbidden (e.g., \xenon\hspace{-3pt}). $$ \prescript{136}{54}{\text{Xe}} \longrightarrow \prescript{136}{56}{\text{Ba}}^{+2} + 2e^{-} + 2\Bar{\nu}_e.$$
If neutrinos are their own antiparticles, it could be possible to observe an event where the anti-neutrinos from \tuvbb annihilate each other, resulting in a `neutrinoless' double-beta decay. 

Over the past decade, several experiments have been searching for \ovbb in different isotopes, using different techniques, yet no discovery has been made to date. Current sensitivity limits of the \ovbb half-life reach $10^{25}-10^{26}$ years \cite{gerda,cuore,kamland2022}. Next-generation experiments are being developed with increased sensitivity to \ovbb by one to two orders of magnitude.

nEXO is a proposed tonne-scale experiment that seeks to search for \ovbb in the isotope \xenon with the goal to reach a sensitivity beyond $10^{28}$ years \cite{nexo_design}. The nEXO detector is a single-phase  Time Projection Chamber (TPC), which will hold 5 tonnes of liquid xenon (LXe) isotopically enriched in the \tuvbb isotope \xenon to 90\%. TPCs are powerful detectors that allow the reconstruction of the position, energy, and topology of each event \cite{exo-200}. This opens the door to locate events in the detector volume and then extract from it and identify the\xenon decay daughter \barium\hspace{-3pt}. This technique is called Ba-tagging and is being developed as a future upgrade to nEXO.
A successful Ba-tagging technique will allow unambiguous identification of a decay of interest as a true double-beta decay event, virtually eliminating all gamma background events.

 \section{Ba-tagging scheme}
 \label{batag}

The technique for detecting \barium in a Xe-based detector using laser spectroscopy was first proposed by M. Moe to facilitate the search for \ovbb\hspace{-3pt} \cite{original_battaging}. Since then, identification of Ba has been demonstrated in neutral Ba \cite{batag2}, in Ba$^+$ \cite{laser_spec_old, laser_spec}, and in Ba$^{2+}$ \cite{next}

A remaining challenge with implementing Ba-tagging is the extraction and isolation of a single \bariumium from tonnes of LXe. Several groups within the nEXO collaboration are working on various approaches to Ba-tagging by combining ion trapping and spectroscopic techniques \cite{batag2,next,batag1,batag3,batag4}.

One approach is based on well-established ion-manipulation techniques to extract and identify the ion, and it is schematically presented in Figure \ref{fig:batag}.

The \bariumium extraction process is initiated when a \ovbb\hspace{-3pt}-like event is detected inside the TPC. 
A capillary is inserted into the detector volume, and its opening is moved close to the reconstructed location where the \bariumium is expected. The \bariumium will be flushed out of the detector volume by the LXe through the capillary. The LXe undergoes a phase transition to gas as it travels through the capillary, leaving the \bariumium in a high-pressure xenon environment. An RF carpet is proposed to guide the ion from the capillary to a converging-diverging nozzle that injects the ion into an RF-only ion funnel \cite{batag1}. There, RF potentials separate the ion from electrically neutral xenon, transporting the \bariumium from a high-pressure environment (2-3 bar) to a high-vacuum level ($<10^{-6}$ mbar). The ion is then injected into a quadrupole mass filter (QMF) to suppress possible accompanying background ions. After filtering, the ion enters a linear Paul trap (LPT), where first it is collisionally cooled with helium gas \cite{buffer} and then trapped. Laser fluorescence spectroscopy of the trapped ion confirms the presence of Ba$^+$ \cite{laser_spec,laser_spec_thesis}. After elemental identification, the ion is injected into a multiple reflection time-of-flight mass spectrometer (MRTOF) to determine its mass and confirm the Ba$^+$ ion to be $\prescript{136}{}{\text{Ba}^+}$. 

Performance of the RF funnel was first studied in \cite{batag1} where ions were produced by a $\prescript{252}{}{\text{Cf}}$ fission source placed inside the Xe just upstream of the RF funnel assembly. Further testing of the RF funnel, which was done using a linear quadrupole ion trap \cite{funnel_thesis}, found evidence of background ions and charged molecular entities like xenon and xenon dimers. 

To effectively demonstrate and characterize the extraction process from Xe gas to vacuum, with subsequent ion identification, we have upgraded the RF-funnel assembly with a linear Paul trap, an MRTOF, and an in-gas laser ablation ion source to produce ions in high-pressure (up to 10 bar Xe and up to 7.8 bar Ar) gas environments \cite{melissa}.

\section{Characterization of ion-extraction from xenon gas}
\label{gas_extract}

RF Funnel, QMF, LPT, and MRTOF for ion extraction from gaseous xenon are currently being assembled and commissioned at McGill University. 

\begin{figure}[h]
    \centering
    \includegraphics[width=0.375\textwidth]{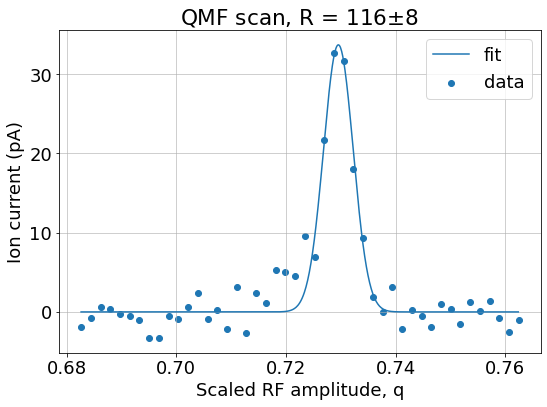}
    \caption{Mass peak of Cs-133 ion obtained from scanning potentials of QMF. The data is fit with a gaussian distribution and resolving power estimated to be $R = m/\Delta m = q/\Delta q = 116\pm8$.}
    \label{fig:qmf}
\end{figure}

The quadrupole mass filter (QMF) is located upstream from LPT and filters incoming ions based on their mass-to-charge ratio. Initial tests performed using a Cs thermal ion source yielded a resolving power, $R = m/\Delta m > 100$, as shown in Figure \ref{fig:qmf}, sufficient for filtering most background ions such as xenon dimer, xenon hydride-hydroxide, krypton-xenon excimers but not the isobar $\prescript{136}{}{\text{Xe}^+}$ \cite{yang}. Currently, the LPT is at the final stages of commissioning, with ion trapping and bunching remaining to be optimized. The electrodes of the Paul trap are strategically designed to optimize clear visibility and accessibility to the trap center. Once ion cooling and bunching have been established, the system can be modified to enable laser fluorescence spectroscopy.

The MRTOF \cite{kevin_paper} was commissioned in parallel to the LPT and is currently fully operational, using ions generated by a laser ablation source \cite{las_manuscript}. With copper ion bunches created by laser ablation, the maximum resolving power achieved using the two mass peaks from naturally occurring isotopes of copper ($\prescript{63}{}{\text{Cu}}$, $\prescript{63}{}{\text{Cu}}$) was determined to be 20,000. After a successful demonstration of ion bunching with the LPT, the MRTOF will be operated using bunched ions, at which point the resolving power is expected to reach 100,000 \cite{titan_mrtof}.

\section{Summary and Future outlook}
\label{future}

Ba-tagging is an active background rejection technique unique to \ovbb searches in Xe-based detectors like nEXO. Several groups are developing Ba-tagging techniques for a possible future upgrade to nEXO. In addition, the group at TRIUMF is developing an accelerator-based Ba-ion source where radioactive $\prescript{139}{}{\text{Cs}}$ ($T_{1/2}$ = 9.3 mins) is injected into a liquid Xe volume instrumented as a TPC, where subsequent decays to $\prescript{139}{}{\text{Ba}}$ are identified, and localized \cite{batag_proposal}. These ions will then be extracted and identified with the aim of determining the overall efficiency of the Ba-tagging technique.

In the very near future, ion bunching of LPT will be demonstrated using a downstream channeltron, and subsequently, MRTOF operation will be optimized using bunched ions from the LPT. Following the completion of commissioning, ions produced in $\prescript{252}{}{\text{Cf}}$ fission will be extracted from Xe gas via the RF funnel and subsequently identified by the combined LPT and MRTOF assembly. Future work will then focus on developing an RF carpet to transport Ba-ions from an in-gas laser ablation ion source in high-pressure Xe gas to a nozzle where they will be injected into the RF funnel. This setup will then be coupled to the capillarity, and its ion extraction efficiency will be quantified using the aforementioned accelerator-driven Ba-ion source. Following the successful demonstration of the entire Ba-tagging scheme, this technique will be considered for the future upgrade to nEXO to further push the detector's sensitivity.

\section*{Acknowledgement}
\label{thanks}

We would like to extend our appreciation to the machine shop at Universit\'e de Montr\'eal for their expert machining of several parts for the Ba-tagging system at McGill University. We would like to express our gratitude to Eamon Egan for their invaluable contribution to the development of the RF system for the LPT. Eamon's expertise in debugging electronic issues was particularly instrumental at various stages of the commissioning process. We are grateful to all our nEXO collaborators for their unwavering support and guidance throughout this project. The work has been supported by the Canada Foundation for Innovation (CFI), the Natural Sciences and Engineering Research Council of Canada (NSERC), the McDonald Institute, and McGill University. The development of the RF funnel at Stanford, and associated hardware was supported by the US National Science Foundation.


\bibliographystyle{elsarticle-num} 
\bibliography{reference}

\end{document}